\documentclass[letterpaper]{article}
\usepackage{aaai16}
\usepackage{times}
\usepackage{helvet}
\usepackage{courier}
\usepackage{multirow}
\usepackage{graphicx}
\usepackage{amsmath}
\usepackage{bm}
\usepackage{verbatim}
\usepackage{enumerate}
\usepackage{color,microtype,hyphenat,balance}
\usepackage{tikz}
\usepackage{paralist}
\usepackage{graphicx}
\usepackage{url}
\usepackage{amsfonts}
\usepackage{mathrsfs}
\usepackage{algorithm}
\usepackage{algorithmic}
\usepackage{amsbsy}
\usepackage{mwe}
\usepackage{caption}
\usepackage{subcaption}
\usepackage{colortbl}
\usepackage{float}
\usepackage{stfloats}
\newcommand{\doc}[1]{\textcolor{black}{#1}}

\begin{document}
\title{Tracking Idea Flows between Social Groups}
\author{540}
\author{Yangxin Zhong$^1$, Shixia Liu$^1$\thanks{S. Liu is the
corresponding author.}, Xiting Wang$^1$, Jiannan Xiao$^1$ \and Yangqiu Song$^2$\\
$^1$School of Software, Tsinghua University, Beijing, P.R. China\\
$^2$Lane Department of Computer Science and Electrical Engineering, West Virginia University, United States\\
$^1$\{zhongyx12,wang-xt11,xjn11\}@mails.tsinghua.edu.cn, $^1$shixia@tsinghua.edu.cn, $^2$yangqiu.song@mail.wvu.edu
}
\maketitle

\begin{abstract}
In many applications, ideas that are described by a set of words often flow between different groups.
To facilitate users in analyzing the flow, we present a method to model the flow behaviors
that aims at identifying the lead-lag relationships between word clusters of different user groups.
In particular, an improved Bayesian conditional cointegration based on dynamic time warping is employed to learn links between words in different groups.
A tensor-based technique is developed to cluster these linked words into different clusters (ideas) and track the flow of ideas.
The main feature of the tensor representation is that we introduce two additional dimensions to represent both time and lead-lag relationships.
Experiments on both synthetic and real \doc{datasets} show that our method is more effective than methods based on traditional clustering techniques and achieves better accuracy.
A case study \doc{was} conducted to demonstrate the usefulness of our
method in helping users understand the flow of ideas between different user groups on social media.
\end{abstract}

\section{Introduction}
As stated in Webster's Third Edition, an idea is ``a formulated thought or opinion.''
Hundreds of millions of users post their ideas on bursty events, hot topics, and personal lives on social media.
Users from different social groups tend to interact with each other on different ideas.
For example,
Democrats and Republicans often interact and communicate on ideas of interest on Twitter.
One idea is about the bipartisan senate immigration bill, which can be described \doc{as} ``immigration, bill, act, law, bipartisan.''
As in this example, a set of words is often used to represent an idea~\cite{Blei2006}.
Frequently, idea flows between different user groups on social media to reflect social interaction and influence~\cite{Pentland2014book}.
These flows facilitate the transfer of information, opinions, and thoughts from group to group.
For instance, idea flows between Democrats and Republicans disclose their leadership and impact on different issues ranging from presidential approval \doc{ratings and} immigration \doc{to health care and} women's equality.
For the immigration idea, Democrats led the discussion of the idea since they called \doc{for} this new immigration bill.
However, most Republican voters \doc{were} concerned that an immigration bill might not solve border security problems.
As they argued and debated with each other, the immigration idea \doc{flowed} between Democrats and Republicans.
For more details about this example, please refer to our case study in the evaluation section.
In many applications, it is desirable to track such \doc{an} idea flow and its lead-lag relationships between different groups~\cite{liu2015exploring,wu2013lead}.

For this reason, the study of idea (influence) propagation has received a great deal of attention~\cite{Leskovec2009,Myers2012,Shaparenko2007,wu2014opinionflow,yu2015micro}.  
\doc{Existing methods} range from link-based \doc{and} content-based to a hybrid \doc{approach}~\cite{Nallapati2008-1}.
Although these methods have been successful in analyzing individual users and their connections, as well as the paths the information \doc{propagates} through,
less attention has been \doc{paid} to studying how the ideas correlate  social groups and interact with each other along time.

The goal of our work is to identify an idea as a word cluster and track the lead-lag relationships between word clusters of different user groups.
To achieve this goal, we first derive an augmented bipartite word graph based on the correlations and lead-lag relationships between words.
Each word is represented by a time series, which encodes its term frequency change over time.
Since the correlation between two words can be irregular and arbitrary \doc{over} time, we provide a way to automatically identify the time period \doc{in which} two words are correlated.
Specifically, we use dynamic time warping (DTW) to align two time series under the monotonic and slope constraint conditions of different time points~\cite{Sakoe1978}.
Then, we employ Bayesian conditional cointegration (BCC)~\cite{Bracegirdle2012} to discover the local correlation between two time series.
After applying BCC, we \doc{can determine} whether two words have a lead or lag relationship at a \doc{particular} time point.
Consequently, we formulate the augmented bipartite graph as a tensor representation.
\doc{In contrast to} traditional time dependent data analysis using tensor, which \doc{employs} one additional dimension to represent time information~\cite{Sun2006,Sun2008}, we \doc{have introduced} two additional dimensions to represent both time and lead-lag relationships.
Moreover, we automatically discover ideas that are represented by clusters of words by factorizing the 4-order tensor.
For a certain pair of ideas, we further apply tensor factorization to cluster the time points, which can be used to segment the time series to identify the lead and lag period between ideas.

To demonstrate the effectiveness of our approach, two experiments and a case study were conducted.
First, we \doc{used} a synthetic dataset, in which the ground truth about word clusters and their lead-lag information is known, to evaluate how the algorithm performs with different noise levels.
Second, we \doc{used} ten time series benchmark datasets to \doc{show} that our algorithm \doc{improves} the time series clustering quality by introducing local lead-lag information.
Third, a case study \doc{was} conducted based on a set of tweets
posted by 514 members of the 113th U.S. Congress in 2013.
\doc{The aim is}
to demonstrate how the ideas from different social groups interact with each other.

\section{Algorithm Overview}

\begin{figure}[t]
	\centering
	\includegraphics[width=0.48\textwidth]{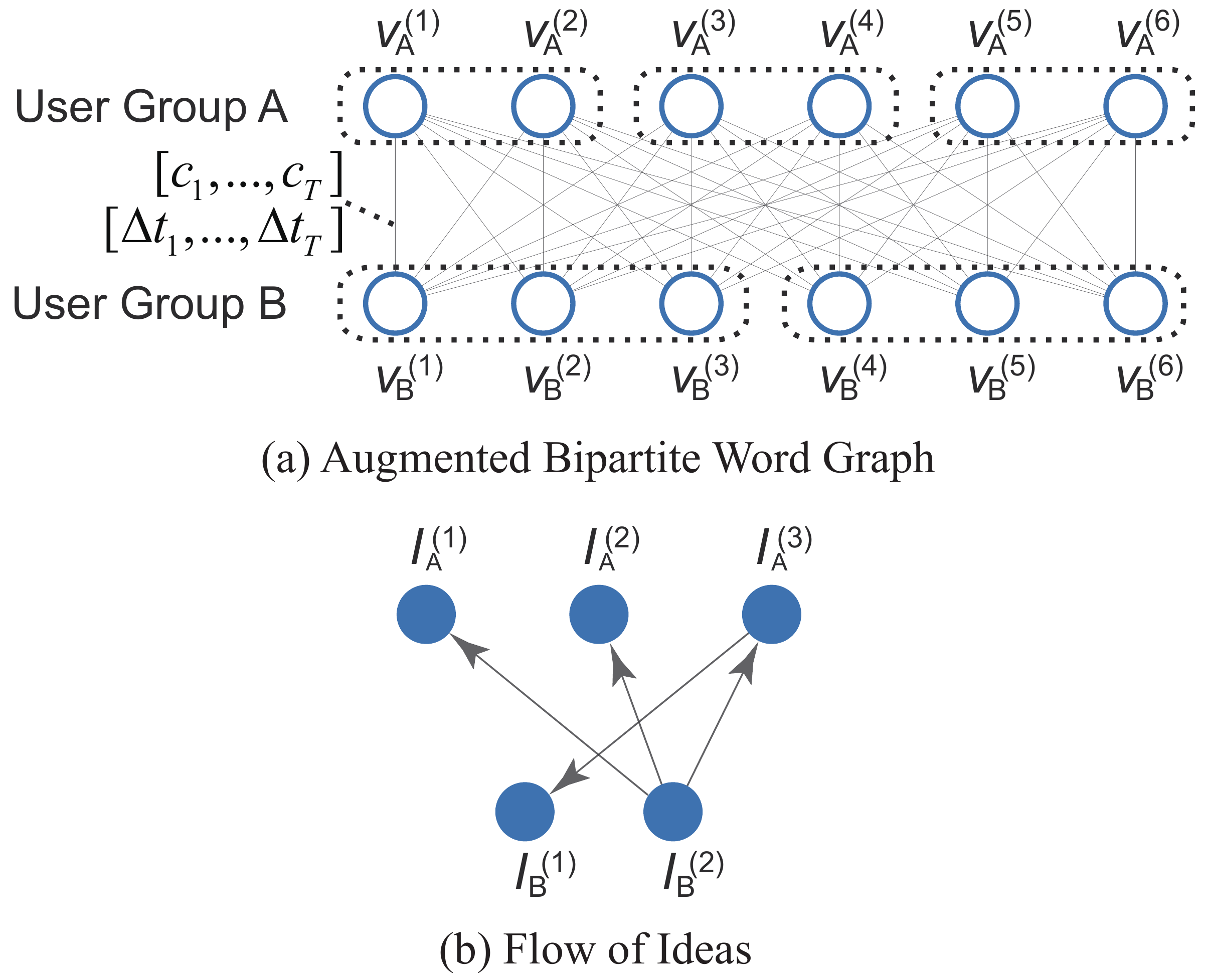}
	\caption{Basic idea of the idea flow model: (a) each edge is represented by a correlation vector $\mathbf{c}=\left[c_1,..., c_T\right]$ and a lead-lag vector ${\bf \Delta t}=\left[\Delta t_1,...,t_T\right]$.
$T$ is the number of time points and $c_k$, $\Delta t_k$ are the correlation value and lead-lag time between two words for the $k$th time point;
(b) words are aggregated into ideas and edges are aggregated into flows. }\label{fig:formulation}
\end{figure}

Suppose each word is tracked by a time series that encodes its frequency \doc{change} over time.
The lead and lag relationships between ideas of different user groups can be derived \doc{using} temporal correlations between words.
As shown in Fig.~\ref{fig:formulation}, we first extract the correlation and lead-lag relationship between words.
An augmented bipartite graph $G = (V, E)$ is used to encode all the correlations and lead-lag relationships between two groups of words.
Second, we discover ideas that are represented by clusters of words as well as their flows between different user groups.
The steps are detailed as below.

\begin{figure}[t]
	\centering
	\includegraphics[width=0.48\textwidth]{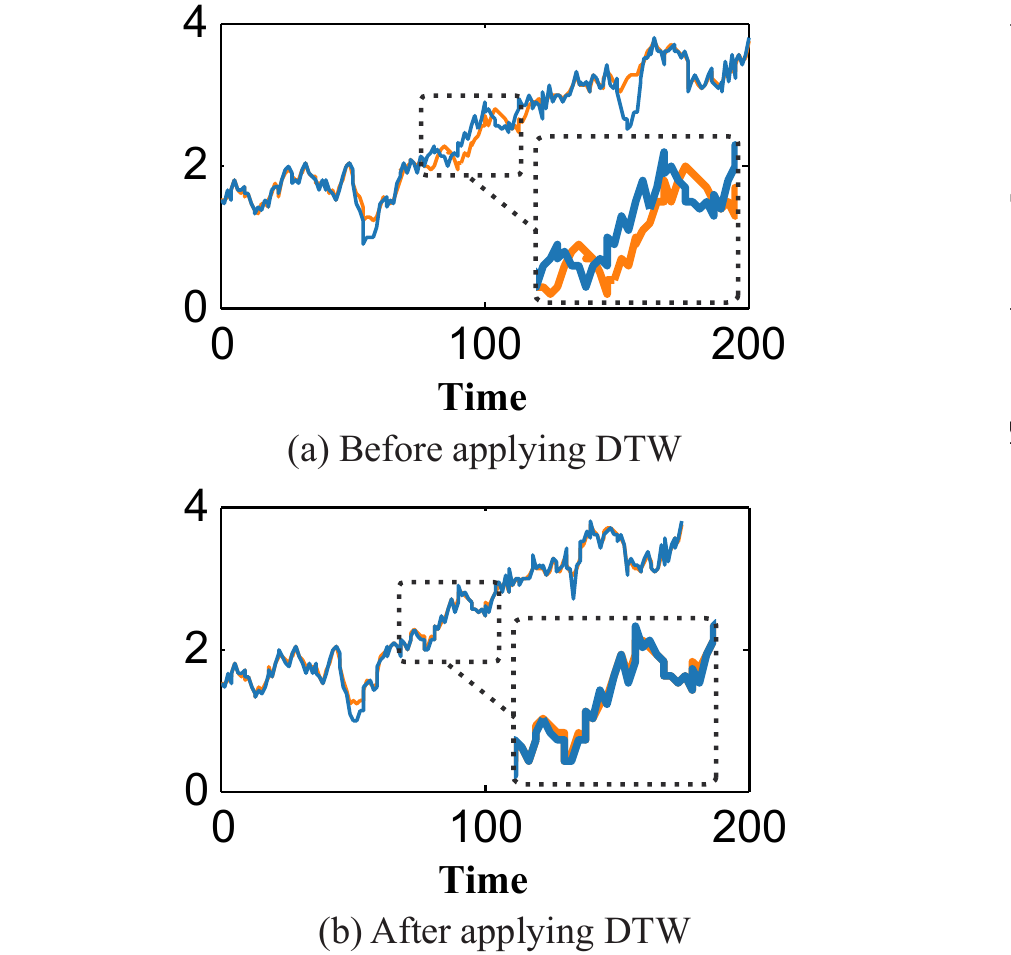}
	\caption{An example of applying the DTW alignment.}\label{fig:model2}
\end{figure}

\begin{compactitem}
\item \textbf{Augmented bipartite graph construction.}
In the first step, we calculate correlations between words to construct the augmented bipartite graph.
Generally, correlations between words can be irregular and arbitrary \doc{over} time \doc{for} two reasons:
1) the correlation of two words may change over time;
2) there are lead-lag relationships between the correlated parts (Fig. ~\ref{fig:model2}(a)).
BCC is able to identify time periods \doc{in which} two time series \doc{are} correlated, but it is not able to detect the lead-lag relationships.
To solve this problem, we developed an improved BCC algorithm that incorporates DTW to align two time series and detect the lead-lag relationships.
With these correlations and lead-lag relationships, an augmented bipartite graph is built,
in which $V = V_A \cup V_B$ represents the words that belong to user groups $A$ or $B$.
Each edge in $E$ is represented by two vectors:
1) correlation vector $\mathbf{c}=\left[c_1,..., c_T\right]$, where $T$ is the number of time points and $c_k=1$ ($c_k=0$) means $v_A^{(i)}$ is correlated (not correlated) with $v_B^{(j)}$ at the $k$th time point;
2) lead-lag vector ${\bf \Delta t}=\left[\Delta t_1,  ..., \Delta t_T\right]$, where $\Delta t_k$ is the lead-lag time between words at time $k$.

\item \textbf{Tracking idea flows.}
In this step, ideas are derived by partitioning the augmented bipartite graph into word clusters.
The key challenge is to partition the augmented bipartite graph, in which each edge is represented by two vectors instead of a real number.
To solve this problem, we model the augmented bipartite graph as a tensor, which uses additional dimensions to represent time and lead-lag relationships.
We then apply tensor factorization techniques to extract a feature vector for each word and cluster the words based on these feature vectors.
According to the clustering results, words are aggregated into ideas and edges are aggregated into flows (Fig.~\ref{fig:formulation}(b)).
For each pair of ideas, we further identify their lead and lag period by clustering the time points using tensor-based techniques.\looseness=-1
\end{compactitem}

\section{Augmented Bipartite Graph Construction}



\doc{This section introduces} our algorithm for calculating correlations and lead-lag relationships between words of different groups by combining BCC with DTW.
Based on these correlations, we then derive an augmented bipartite word graph.
Suppose we have two time series $x_{1:T}$ and $y_{1:T}$ for two words that belong to different user groups.
The goal is to extract the correlation vector $\mathbf{c}=\left[c_1,..., c_T\right]$ and the lead-lag vector ${\bf \Delta t}=\left[\Delta t_1,  ..., \Delta t_T\right]$.
In particular, our algorithm consists of the following two steps.

First, DTW is employed to calculate $\left[\Delta t_1,  ..., \Delta t_T\right]$.
DTW is a widely used dynamic programming algorithm that aligns two time series under the monotonic and slope constraint conditions of different time points~\cite{Sakoe1978}.
We choose DTW because it is both flexible and efficient~\cite{Sakoe1978}.
An example of applying DTW is shown in Fig.~\ref{fig:model2}.
Before applying DTW, there are misalignments between correlated parts of $x_{1:T}$ and $y_{1:T}$ (Fig.~\ref{fig:model2}(a)).
After applying DTW, the misaligned parts are successfully aligned together (Fig.~\ref{fig:model2}(b)).
$\Delta t_k$ is calculated based on this alignment: if $x_k$ is aligned to $y_l$, we set $\Delta t_k$ to $l-k$.\looseness=-1

Second, we derive $\left[c_1,..., c_T\right]$ by using BCC to examine the cointegration between aligned time series $x'_{1:T}$ and $y'_{1:T}$.
Cointegration describes a relationship between time series where there \doc{is} a stationary linear combination.
Specifically, if $x'_k$ and $y'_k$ are cointegrated in a given time period, we have\looseness=-1
\[y'_k = \alpha + \beta x'_k + \epsilon_k,\]
where $\alpha$ is a constant and $\beta$ is the linear regression co-efficient.
$[\epsilon_1, \epsilon_2, ... ,\epsilon_T]$ follows a mean-reverting, stationary process.
Compared to typical measures such as \doc{the} Pearson correlation and Spearman correlation~\cite{Bluman2012}, BCC has two advantages~\cite{Bracegirdle2012}.\looseness=-1
\begin{compactitem}
\item BCC is able to calculate both global and local correlations while Pearson correlation and Spearman correlation can only detect global correlations.
\item BCC produces less spurious results compared with the classical cointegration testing approach.\looseness=-1
\end{compactitem}
After applying BCC, we detect the correlation (cointegration) of $x'_{1:T}$ and $y'_{1:T}$ at different times (${\bf c'}$).
We then assign ${\bf c'}$ to ${\bf c}$ according to the alignment relationships derived by DTW.

\section{Tracking Idea Flows}

This section introduces the tensor representation as well as tensor-based augmented bipartite graph partition and aggregation.\looseness=-1

\subsection{Tensor Representation}

\begin{figure}[t]
	\includegraphics[width=0.48\textwidth]{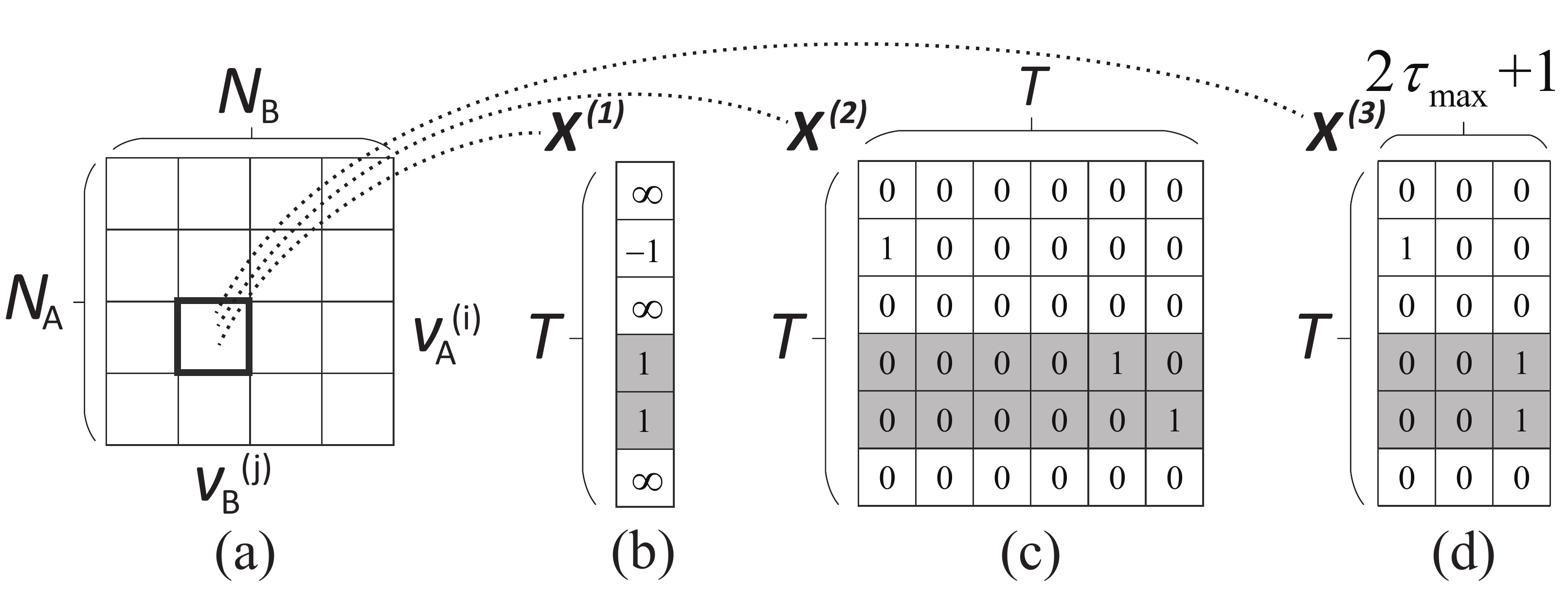}
	\caption{Three tensor representations: (a) word matrix; (b) (c) (d) three representations to encode time and lead-lag relationships.}\label{fig:tensordesign}
\end{figure}

The key challenge of partitioning the augmented bipartite word graph is that each edge is represented by two vectors rather than a real number (Fig.~\ref{fig:formulation}(a)).
To tackle this challenge, we model the augmented bipartite word graph as a tensor, which is able to encode time and lead-lag relationships using additional dimensions.

A straightforward tensor representation is ${\bf X}^{(1)}\in$  $\mathbb{R}^{N_A \times N_B \times T}$ (Fig.~\ref{fig:tensordesign}(b)).
Here $N_A$ and $N_B$ are the numbers of words in user groups $A$ and $B$, respectively.
In our implementation, user groups are already identified by two professors who majors in media and communications.
${\bf X}^{(1)}_{ijk}$ represents the lead-lag relationship between the $i$th word in $A$ and the $j$th word in $B$ at the $k$th time point.
We set ${\bf X}^{(1)}_{ijk}$ to $\Delta t_k$ if the two words are correlated at the $k$th time point and $\infty$ if the two words are not correlated at that time point.
However, representing uncorrelated information as $\infty$ may not be good enough because it can also mean \doc{one word} correlated with another with \doc{an infinite} time lead.
Moreover, ${\bf X}^{(1)}$ is very dense.
As a result, computation based on ${\bf X}^{(1)}$ is very expensive.\looseness=-1

\begin{figure*}[bp]

   \begin{minipage}{\textwidth}
      \centering
      \subcaptionbox{Flow\_F1 vs. Noise Level.\label{fig:exp1-FlowF1}}
      {\includegraphics[width=0.32\textwidth]{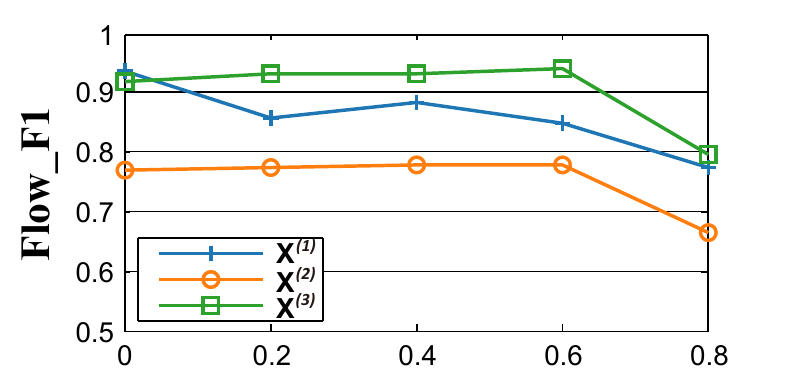}}\quad
      \subcaptionbox{FlowLead\_F1 vs. Noise Level.\label{fig:exp1-FlowLeadF1}}
       {\includegraphics[width=0.32\textwidth]{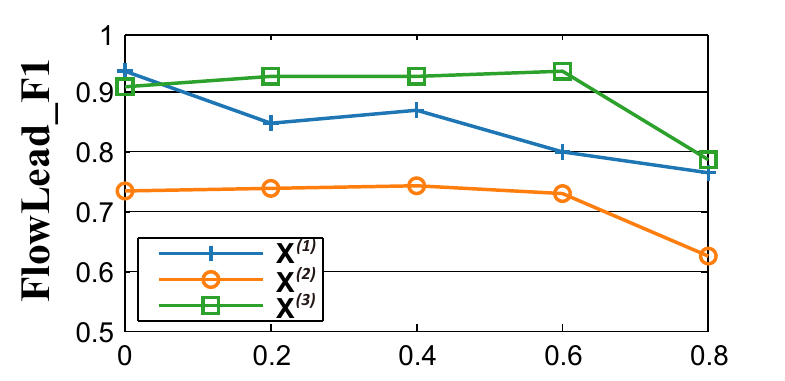}}
       \subcaptionbox{FlowLeadTime\_F1 vs. Noise Level.\label{fig:exp1-FlowDelayTimeF1}}
       {\includegraphics[width=0.32\textwidth]{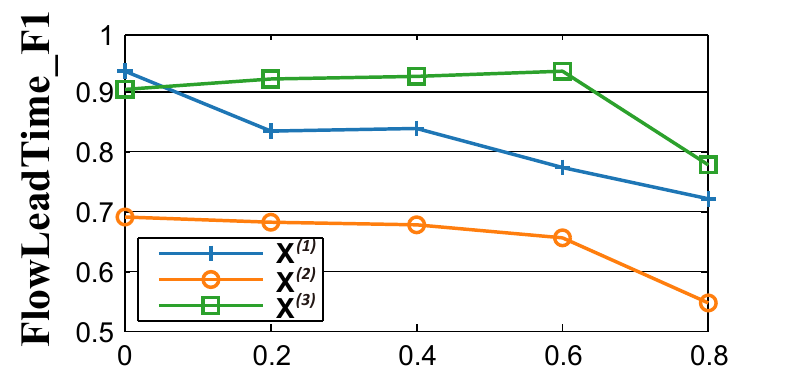}}
    \end{minipage}
    \begin{minipage}{\textwidth}
    \centering
    \subcaptionbox{Lead-lag Time MSE vs. Noise Level.\label{fig:exp1-MSE}}
    {\includegraphics[width=0.32\textwidth]{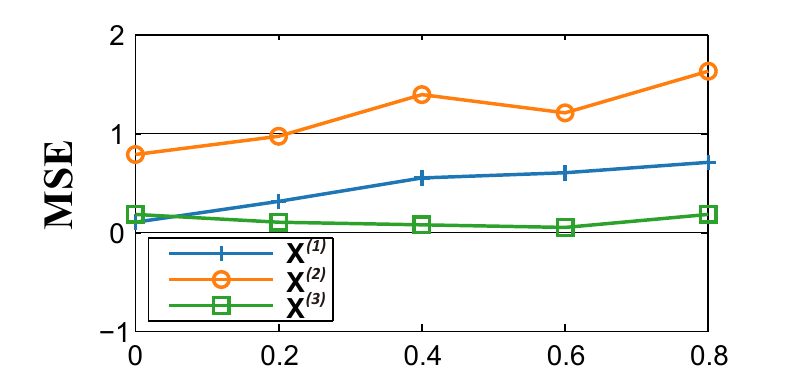}}\quad
    \subcaptionbox{Word Clustering NMI vs. Noise Level.\label{fig:exp1-NMI}}
     {\includegraphics[width=0.32\textwidth]{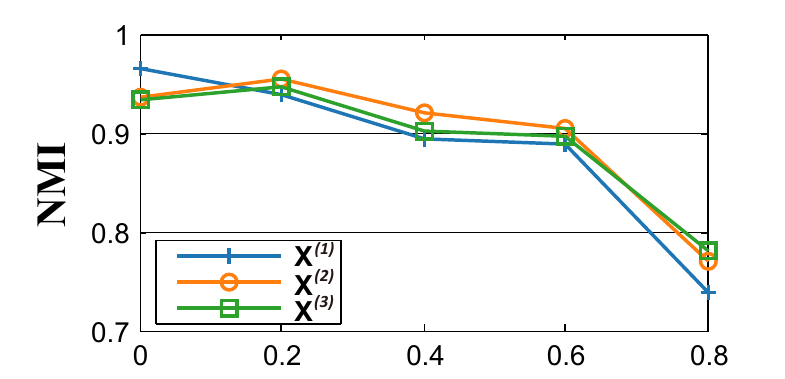}}
     \subcaptionbox{Running Time vs. Noise Level.\label{fig:exp1-Time}}
     {\includegraphics[width=0.32\textwidth]{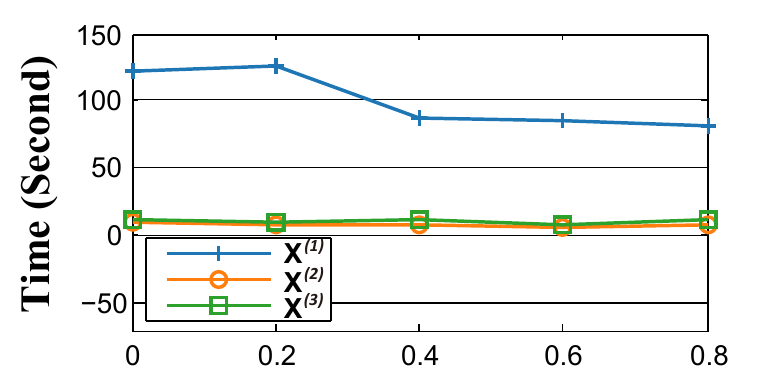}}
  \end{minipage}
\caption{Comparison of three tensors (${\bf X}^{(1)}$, ${\bf X}^{(2)}$, and ${\bf X}^{(3)}$) in terms of accuracy and efficiency at different noise levels.}\label{fig:exp1}
\end{figure*}



A more reasonable solution is to extend the 3-order tensor to a 4-order representation: ${\bf X}^{(2)}\in  \mathbb{R}^{N_A\times N_B\times T\times T}$ (Fig.~\ref{fig:tensordesign}(c)).
${\bf X}^{(2)}_{ijkl}$ is set to 1 if and only if the $i$th word in user group $A$ is correlated to the $j$th word in user group $B$ at the $k$th time point and $l=k+\Delta t_k$.
Otherwise, ${\bf X}^{(2)}_{ijkl}$ is set to 0.
Compared to ${\bf X}^{(1)}$, ${\bf X}^{(2)}$ is sparse and it clearly distinguishes \doc{between} uncorrelated time points \doc{and} correlated time points.
However, ${\bf X}^{(2)}$ fails to represent lead-lag time effectively.
As shown in Fig.~\ref{fig:tensordesign}, for words $v_A^{(i)}$ and $v_B^{(j)}$, the lead-lag time of the 4th time point and 5th time point are both 1.
However, the feature vectors of ${\bf X}^{(2)}$ at the 4th time point and the 5th time point (grey rows) are different ($[0,0,0,0,1,0]$ and $[0,0,0,0,0,1]$).
This leads to incorrect lead-lag results along the time dimension.


To solve this issue, we redesign the $4$-th dimension of ${\bf X}^{(2)}$ to more effectively encode lead-lag time.
Our design is ${\bf X}^{(3)}\in  \mathbb{R}^{N_A\times N_B\times T\times (2\tau_{max}+1)}$ (Fig.~\ref{fig:tensordesign}(d)),
where $\tau_{max}$ is the maximum allowable time deviation between two aligned time points in DTW.
${\bf X}^{(3)}_{ijkl}$ is set to 1 if and only if the $i$th word in $A$ and the $j$th word in $B$ are correlated at the $k$th time point and $l=\Delta t_k+\tau_{max}+1$.
Otherwise, ${\bf X}^{(3)}_{ijkl}$ is set to 0.
As shown in Fig.~\ref{fig:tensordesign}(d), the feature vectors of ${\bf X}^{(3)}$ at 4th and 5th time points are the same ($[0,0,1]$).
As a result, the 4th time point and 5th time point will be clustered into one segment and we can accurately detect the lead and lag periods.\looseness=-1



\subsection{Augmented Bipartite Graph Partition}

Based on the tensor representation, our algorithm first employs tensor-based techniques to extract the feature vector for each word and then clusters the words using these features vectors.
The algorithm consists of three steps.

First, we employ a tensor factorization algorithm called greedy PARAFAC~\cite{Kolda2005} to factorize the tensor.
This algorithm is adopted because it is efficient and is able to deal with tensors that have more than three dimensions.
The algorithm yields a rank-$q$ approximation of ${\bf X}^{(3)}$ in the form
\[{\bf X}^{(3)} \approx \sum_{m = 1}^{q} \lambda^{(m)} \bm{u}^{(m)} \circ \bm{v}^{(m)} \circ \bm{w}^{(m)}\circ \bm{h}^{(m)},\]
where $\lambda^{(m)} \in \mathbb{R}$, $\bm{u}^{(m)} \in \mathbb{R}^{N_A}$,  $\bm{v}^{(m)} \in \mathbb{R}^{N_B}$, $\bm{w}^{(m)} \in \mathbb{R}^T$, $\bm{h}^{(m)} \in \mathbb{R}^{(2\tau_{max}+1)}$.
$\bm{u}^{(m)} \circ \bm{v}^{(m)} \circ \bm{w}^{(m)} \circ \bm {h}^{(m)}$ is the 4-way outer product so that $(\bm{u}^{(m)} \circ \bm{v}^{(m)} \circ \bm{w}^{(m)} \circ \bm {h}^{(m)})_{ijkl} = \bm{u}^{(m)}_i \bm{v}^{(m)}_j \bm{w}^{(m)}_k \bm {h}^{(m)}_l$.
Here $\bm{u}^{(m)}_i$ is the $i$th entry of vector $\bm{u}^{(m)}$.

Then we use factors $\{\bm{u}^{(m)}\}$ and $\{\bm{v}^{(m)}\}$ to extract feature vectors for each word in groups $A$ and $B$, respectively.
Take user group $A$ as an example.
The feature matrix $\bm{U}$ for this group is constructed as follows:
\[\bm{U} = [\bm{u}^{(1)}, \bm{u}^{(2)}, ... , \bm{u}^{(q)}] \in \mathbb{R}^{N_A \times q}.\]
For the $i$th word in user group $A$, the feature vector is set to $[{\bm U}_{i1},...,{\bm U}_{iq}]$. 

Finally, we utilize a clustering algorithm such as K-means, on the extracted feature vectors to detect word clusters.

\subsection{Augmented Bipartite Graph Aggregation}

Based on the clustering results, we aggregate words into ideas and edges into flows (Fig.~\ref{fig:formulation}).
To identify the lead and lag periods between ideas, we segment the time series for each idea pair by further applying tensor factorization.
Take ideas $I_A^{(1)}$ and $I_B^{(2)}$ as an example.
First, we extract a sub-tensor from ${\bf X}^{(3)}$ that belongs to $I_A^{(1)}$ and $I_B^{(2)}$.
Without a loss of generality, we assume that the $1$st to $N_1$th words in $A$ belong to $I_A^{(1)}$ and the $1$st to $N_2$th words in $B$ belong to $I_B^{(2)}$.
Then the sub-tensor is denoted as ${\bf X}^{(3)}_{1:N_1, 1:N_2, 1:T, 1:(2\tau_{max}+1)} \in \mathbb{R}^{N_{1}\times N_{2}\times T\times (2\tau_{max}+1)}$.
After extracting the sub-tensor, we apply tensor factorization on it to derive a feature vector for each time point and cluster these feature vectors by using a clustering algorithm such as K-means.
Based on the clustering results, we divide the time points into segments.
Finally, we extract the correlation between $I_A^{(1)}$ and $I_B^{(2)}$ at the $k$th segment ($\bar c_k$).
We average $c_k$ for all word pairs at time points that belong to this segment.
If this value is greater than a given threshold, $\bar c_k=1$; otherwise, $\bar c_k=0$.
If $\bar c_k=1$, the lead-lag time between these two ideas, $\overline{\Delta t}_k$, is computed by averaging $\Delta t_k$.\looseness=-1
\begin{table*}[bp]
\small
\centering
\begin{tabular}{|c|c|c|c|c|c|c|c|c|c|c|}
\hline
                  & Coffee & L7 & Trace & Fish & OSU Leaf & SC & Strawberry & FA & CC & Wafer
                   \\
\hline
$K$        &    2 &  7  & 4 & 7 & 6 &  6 & 2 & 14  & 3& 2 \\
\hline
$N$        & 56   & 143  & 200   &   350  & 442 & 600 & 983 & 2250 & 4307 & 7174 \\ \hline
$T$        & 286   & 319  &  275  & 463 & 427 & 60 & 235 & 131 & 166 & 152 \\ \hline
\end{tabular}
\caption{Summary statistics of the 10 UCR time series datasets.
Here $K$ represents the number of classes, $N$ is the number of time series, and $T$ denotes the number of time points.
The dataset names L7, SC, FA, and CC are abbreviations for Lightning-7, Synthetic Control, Face (All), and Chlorine Concentration, respectively.
}
\label{table:exp2dataset}
\end{table*}

 \begin{table*}[bp]
\small
\centering
\begin{tabular}{|c|p{1.25cm}|p{1.25cm}|p{1.25cm}|p{1.25cm}|p{1.25cm}|p{1.25cm}|p{1.25cm}|p{1.25cm}|p{1.25cm}|p{1.25cm}|}
\hline
                  &  \ \ \ Coffee & \ \ \ \ \ \ L7  &  \ \ \ \ Trace & \ \ \ \ \ Fish & OSU Leaf &\ \ \ \ \ \ SC & Strawberry &  \ \ \ \ \ \ FA  & \ \ \ \ \ \ CC & \ \ \ Wafer  \\
\hline
B1        & 0.00$\pm$0.00 & 0.43$\pm$0.02 & 0.53$\pm$0.02  & 0.29$\pm$0.03 & 0.22$\pm$0.02 & 0.78$\pm$0.03 & 0.12$\pm$0.00 &  0.37$\pm$0.02 & 0.01$\pm$0.00 & 0.00$\pm$0.00  \\
\hline
B2        & 0.54$\pm$0.00  & 0.08$\pm$0.00 & 0.45$\pm$0.00  &  \textbf{0.53$\pm$0.02} & 0.02$\pm$0.00 & 0.59$\pm$0.01 & 0.13$\pm$0.00  &  0.24$\pm$0.00 & 0.01$\pm$0.00 & 0.04$\pm$0.00 \\
\hline
B3        & 0.48$\pm$0.00 & 0.22$\pm$0.03 & 0.54$\pm$0.00 & 0.12$\pm$0.00 & 0.13$\pm$0.00 &   0.46$\pm$0.06 &  \ \ \ \ \ \ \ $\backslash$ & \ \ \ \ \ \ \ $\backslash$  & \ \ \ \ \ \ \ $\backslash$  & \ \ \ \ \ \ \ $\backslash$ \\
\hline
Ours       & \textbf{0.64$\pm$0.00} & \textbf{0.44$\pm$0.02} & \textbf{0.64$\pm$0.00} & 0.16$\pm$0.01  &  \textbf{0.26$\pm$0.00} &   \textbf{0.80$\pm$0.03} & \textbf{0.32$\pm$0.00} &  \textbf{0.39$\pm$0.09} & \textbf{0.02$\pm$0.00} & \textbf{0.19$\pm$0.00}  \\
\hline
\end{tabular}
\caption{
Comparison of our tensor-based clustering method with three baselines in terms of NMI on 10 datasets of the UCR time series benchmark datasets.
}
\label{table:exp2results}
\end{table*}

\section{Evaluation}
\label{sec:evaluation}


We conducted two experiments and a case study to illustrate the effectiveness and usefulness of our idea flow tracking method.
All the experiments were conducted on a workstation with an Intel Xeon E52630 CPU (2.4 GHz) and 64GB of Memory.\looseness=-1


\subsection{Experiment on Synthetic Data}

In this experiment, we used a synthetic dataset to show how our algorithm performs with different noise levels.
We also compared the accuracy and efficiency of three tensor representations.\looseness=-1

\noindent \textbf{Experimental Settings.}
In this experiment, synthetic datasets with five different noise levels were generated (noise level $L\in\{0,0.2,...,0.8\}$).
For each $L$, we generated 50 synthetic datasets and averaged the results of these datasets.
For each dataset, we first generated ideas and the lead-lag relationships between ideas (Fig.~\ref{fig:formulation}(b)), which contain $\bar c_k$ and $\overline{\Delta t}_k$.
Then we generated the augmented bipartite word graph (Fig.~\ref{fig:formulation}(a)) according to noise level $L$. 
$L$ denotes the probability that $c_k=0$ when $\bar c_k=1$.
The larger the $L$, the more difficult \doc{it is} to generate accurate idea flows by using the augmented bipartite word graph.
In the synthetic datasets, the number of ideas in each user group varied from 2 to 6,
the number of words in each idea varied from 10 to 30, the lead-lag time varied from -6 to 6, \doc{and} the number of time points in the lead or lag periods varied from 20 to 40, \doc{all while} $T$ was set to 200.
We applied our algorithm with different tensors (${\bf X}^{(1)}$, ${\bf X}^{(2)}$, and ${\bf X}^{(3)}$) \doc{to} the datasets and reported the accuracy and efficiency.\looseness=-1

\noindent \textbf{Criteria.}
We evaluated the accuracy of our algorithm \doc{based on} five criteria.
To evaluate the accuracy of our algorithm in detecting lead-lag relationships between ideas, we used three F1-measures: \textbf{Flow\_F1}, \textbf{FlowLead\_F1}, and \textbf{FlowLeadTime\_F1}.
Flow\_F1 measures the accuracy of our algorithm in estimating $\bar c_k$.
FlowLead\_F1 is more rigorous than Flow\_F1.
It not only measures the accuracy of $\bar c_k$, but also measures the accuracy of our algorithm in detecting the lead or lag status (i.e., whether $\overline{\Delta t}_k>0$, $\overline{\Delta t}_k=0$, or $\overline{\Delta t}_k<0$).
FlowLeadTime\_F1 is even more rigorous than FlowLead\_F1,
which measures both $\bar c_k$ and $\overline{\Delta t}_k$.
A result with larger F1-measures is better.
In addition to F1-measures, we used Mean Squared Error (\textbf{MSE}) to evaluate the effectiveness of our algorithm in detecting the lead-lag time $\overline{\Delta t}_k$.
Smaller MSE values indicate better quality.
Moreover, a widely used measure for clustering quality, Normalized Mutual Information (\textbf{NMI})~\cite{Strehl2003}, was utilized to measure the accuracy of idea identification.
A larger NMI value indicates more accurate idea identification.\looseness=-1

\noindent \textbf{Results.}
Fig.~\ref{fig:exp1} compares the three tensors in terms of accuracy and efficiency at different noise levels.
The following observations can be made from the results.

\noindent \emph{Accuracy.}
Overall, ${\bf X}^{(3)}$ has the best performance in terms of accuracy. 
This demonstrates that ${\bf X}^{(3)}$ is less sensitive to noises compared to ${\bf X}^{(1)}$ by using two additional dimensions to encode both $c_k$ and $\Delta t_k$.
Although ${\bf X}^{(2)}$ is comparable to ${\bf X}^{(3)}$ in terms of NMI, it is worse in terms of F1-measures and MSE.
This indicates that ${\bf X}^{(2)}$ is accurate in identifying ideas, but is not very accurate in detecting lead-lag relationships between ideas.
This is due to its deficiency in clustering the time points and detecting lead and lag periods.

\noindent \emph{Efficiency.}
As shown in Fig.~\ref{fig:exp1}(f), ${\bf X}^{(1)}$ is much slower than ${\bf X}^{(2)}$ and ${\bf X}^{(3)}$.
This is because ${\bf X}^{(2)}$ and ${\bf X}^{(3)}$ are much sparser than ${\bf X}^{(1)}$.
The computation time for ${\bf X}^{(2)}$ and ${\bf X}^{(3)}$ is comparable.
Both of them are relatively efficient with a maximum computation time of 12 seconds.

Because ${\bf X}^{(3)}$ is both accurate and efficient, we used this tensor representation in the following experiments.



\subsection{Experiment on Real Data}

In this experiment, we use real-world time series benchmark datasets to demonstrate that our algorithm can improve the quality of time series clustering by incorporating local lead-lag information.

\noindent \textbf{Experimental Settings.}
The experiment was conducted on ten real-world time series datasets in the UCR archive~\cite{UCRArchive}, in which the class label of each time series is known.
The ten time series datasets include datasets where the shapes of \doc{the} time series are quite similar as well as datasets where the shapes are different.
Table~\ref{table:exp2dataset} shows the summary statistics of the UCR time series benchmark datasets used in the experiment.
NMI was utilized to measure the clustering quality.
We \doc{performed} a grid search on $\tau_{max}$ ($\{1,2,...,10\}$) and chose the parameter with the largest NMI value ($\tau_{max} = 6$).
To reduce \doc{any} bias caused by initial cluster assignments, we ran each experiment 100 times with different random seeds and reported the mean and standard deviation. \looseness=-1


\begin{figure*}[t]
	\centering
	\includegraphics[width=1\linewidth]{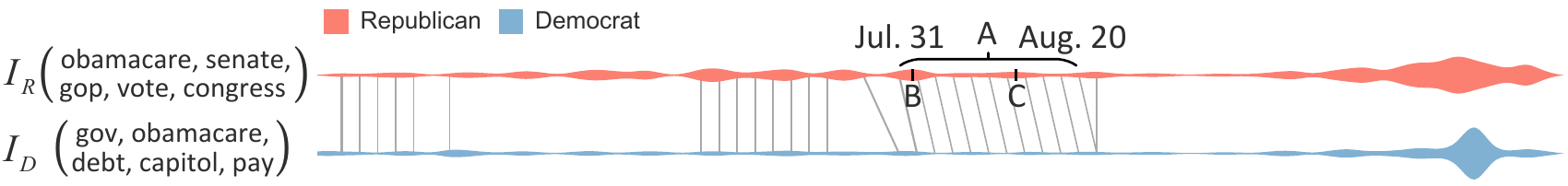}
	\caption{Local lead-lag relationships between ideas in different user groups.
	}\label{fig:casestudyideaflow}
\end{figure*}

\noindent \textbf{Baselines.}
Three baselines were used in this experiment.
The first baseline (\textbf{B1}) \doc{was} K-means~\cite{Hartigan1979} and the second baseline (\textbf{B2}) \doc{was} spectral clustering~\cite{Chen11}.
In B1, we \doc{utilized} $y$-values of time series to form feature vectors.
In B2, we employed $K_{DTW}$ kernel~\cite{Marteau2015} to calculate similarities between time series.
The third baseline (\textbf{B3}) \doc{was} similar to our algorithm.
The difference is that it does not utilize DTW to calculate local lead-lag time.
Instead, it calculates a global lead-lag time by moving the time series forward and backward along the time axis to find the time shift that results in the highest correlation.\looseness=-1

\noindent \textbf{Results.}
Table~\ref{table:exp2results} illustrates the \doc{results} of comparing our tensor-based clustering algorithm with three baselines in terms of NMI.
For B3, several experiments \doc{were unable to} return results within 20 days.
\doc{These} experiments are marked \doc{with a} ``$\backslash$."
As shown in the table, our algorithm performs better than the baselines in \doc{all} datasets except for one (Fish).
The major difference between our algorithm and the baselines is that the baselines did not consider local lead-lag relationships during the clustering process.
In particular,
B1 and B2 did not employ any lead-lag measures and
B3 calculated a global lead-lag time instead of local lead-lag time.
This indicates that the clustering quality can be improved by using local lead-lag relationships.
Next, we examined why our algorithm is worse than B1 and B2 in the Fish dataset.
\doc{We} found time series with similar shapes but different amplitudes are labelled to be different in this dataset.
Since B1 and B2 use absolute $y$-values of the time series, they are more accurate in distinguishing such time series.
Frequently, the time series of words that have similar shapes but different amplitudes are regarded \doc{as} correlated.
As a result, our algorithm is more appropriate to process textual data.

\subsection{Case Study on Twitter Data}

To demonstrate the usefulness of our algorithm in detecting idea flows, we conducted a case study with a professor who majors in media and communications.
The dataset contains 156,501 tweets posted by 514 members of the 113th U.S. Congress from Apr. 9 to Oct. 13, 2013.
The professor and another professor who also majors in media and communications manually divided these members into two groups: 243 Democrats and 271 Republicans.
We segmented the data into 94 time points (every 2 days).
Stopwords and rare words that occur less than 5 times each day by average were removed to reduce noise.
We extracted 15 ideas for each group.\looseness=-1


First, we checked the global lead-lag relationships between all ideas.
We found there \doc{were} 119 lead-lag relationships on pairs of ideas.
Among these pairs, 62 (52\%) were led by Democrats and 57 (48\%) were led by Republicans.
This result demonstrated that Democrats marginally outperformed Republicans in taking the lead at that time.
The professor confirmed our findings.
He said President Obama (Democratic) started to realize the political importance of social media after he was re-elected as President \doc{in} 2012.
After that, Democrats put more effort \doc{into} social media.
Because the president was a Democrat and the Democratic Party \doc{had a} Senate majority, Democrats had \doc{an} advantage over Republicans in leading public opinion.


 \begin{table}[t]
\small
\centering
\begin{tabular}{|c|c|}
\hline
1 & immigration, bill, act, law, bipartisan  \\
\hline
2 & military, join, service, honor, nation   \\
\hline
3 & women, american, succeed, job, equality \\
\hline
\end{tabular}
\caption{Top three ideas led by Democrats.}
\label{table:casestudyideasD}
\end{table}

\doc{We next} examined the ideas that were led by Democrats and Republicans, respectively.
We ranked the ideas by the number of time points \doc{at which} they led other ideas.

The top three ideas for Democrats are shown in Table~\ref{table:casestudyideasD}.
The professor immediately identified that they were \doc{the} bipartisan senate immigration bill, honoring military service members, and women's equality.
By checking related tweets, the professor found Democrats took the lead on these ideas because they were more supportive of them.
Take the ``bipartisan senate immigration bill'' idea as an example.
Because Democrats supported this bill, they kept on posting \doc{the} latest information about this bill and hosted new activities (``House leaders should allow debate \& vote on Bipartisan Senate Immigration Bill. It's not perfect but we need reform. http://t.co/btKDxblbQf").
Republicans lagged because they mainly responded by disapproving \doc{of} Democrats' ideas.
One major concern was that the immigration bill might not solve border security problems (``The immigration bill offers false promises about border security and enforcement measures.").
The professor commented \doc{that} this kind of \doc{interaction is} common between Democrats and Republicans.\looseness=-1

 \begin{table}[t]
\small
\centering
\begin{tabular}{|c|c|}
\hline
 1 & tax, economy, reform, family, icymi  \\
\hline
2 & job, plan, student, school, program   \\
\hline
3 & obamacare, senate, gop, vote, congress  \\
\hline
\end{tabular}
\caption{Top three ideas led by Republicans.}
\label{table:casestudyideasR}
\end{table}

Next, we checked the ideas led by Republicans.
As shown in Table~\ref{table:casestudyideasR}, the top three \doc{were} tax reform, Republican's plan to create jobs, and \doc{health care} vote in Congress ($I_R$).
While Republicans were supportive of tax reform and plans to create jobs, they did not support \doc{Obama's health care plan (Obamacare)}.
To understand why $I_R$ was led by Republicans, we examined the correlated ideas from Democrats.
Among them, the idea with the largest lag-time ($I_D$) was exemplified by words \doc{``gov," ``obamacare," ``debt," ``capitol," and ``pay,"} which \doc{all are related to Obamacare}.
To know why $I_R$ led $I_D$, we further checked the local lead-lag relationships between the two ideas.
As shown in Fig.~\ref{fig:casestudyideaflow},
each idea is represented by a colored stripe \doc{while} the $x$-axis represents time.
The width of a stripe changes over time, encoding \doc{the} temporal ``hotness" of this idea.
The ``hotness" is measured by the number of words that talks about this idea at each time.
The local lead-lag relationships are visualized by the links that connect correlated time points.
Fig.~\ref{fig:casestudyideaflow} shows that the main lead period for idea $I_R$ is from Jul. 31 to Aug. 20 (Fig.~\ref{fig:casestudyideaflow}A).
By checking the two peaks (Fig.~\ref{fig:casestudyideaflow}B and Fig.~\ref{fig:casestudyideaflow}C) on the stripe during this time period, we found Republicans led this idea because they were very active in opposing \doc{the health care plan}.
On Aug. 2 (Fig.~\ref{fig:casestudyideaflow}B), they voted to repeal \doc{Obamacare} for the 40th time.
Around Aug. 13 (Fig.~\ref{fig:casestudyideaflow}C), they posted many tweets \doc{criticizing the fact} that a key consumer protection in \doc{the health care plan} was delayed (``Yet another delay for Obamacare via @CBSNews - Key consumer protection in Obamacare delayed http://t.co/0brYxzjUj7").

\section{Related Work}

Our work is relevant to influence analysis between documents.
Previous \doc{research} on influence analysis can be categorized into three groups: content-based, link-based, and hybrid methods.

Content-based methods use the content of documents \doc{and} their temporal changes to analyze influence~\cite{Cui2011-2,liu2012tiara,Shaparenko2007,zhang2010evolutionary}.
Shaparenko and Joachims assumed the language model of a new document is a mixture of \doc{the} language model of all previously published documents~\cite{Shaparenko2007}.
They adopted a likelihood ratio test to detect the influence between documents.
\doc{However, their} method cannot model influence at the word level.
To learn the influence at the word/phrase level, MemeTracker~\cite{Leskovec2009} \doc{employed} a graph-based method to cluster variants of phrases and derive corresponding threads in the news cycle.
To link two given documents \doc{with} a coherent chronological chain, Shahaf and Guestrin~\cite{Shahaf2010} employed a bipartite graph between words and documents and utilized the graph to model the influence between two documents.
Compared with this work, our method builds an augmented bipartite graph between words used by two user groups by employing the improved BCC technique.
The bipartite link is represented by a set of correlation values and lead-lag times between two words at different times instead of a single weight.
As a result, we can model the influence between groups from a global overview to local temporal details.

Link-based methods use links (e.g., citations or follower-followee relationships) in a network to analyze influence.
PageRank~\cite{Brin1998} and its variations~\cite{Haveliwala2002,Ma2008} are typical examples of such methods.
Researchers also studied the problem of influence maximization in a network~\cite{Feng2014,Kempe2003,Ohsaka2014}.
However, link-based methods need to utilize explicit links in the data to model influence.
Such kind of links may not be available in many real-world datasets such as news articles.

To solve this issue, researchers have developed hybrid methods that use \doc{both} links and content to analyze
influence~\cite{Cui2011,El-Arini2011,GomezRodriguez2010}.
For example, given the topics discussed in a network, Tang et al.~\cite{Tang2009} constructed a factor graph to compute the topic level influence in a large network.
To model influence as well as topics in a corpus, researchers have proposed several topic models that incorporate links between documents ~\cite{Dietz2007,Erosheva2004,Gerrish2010,Guo2010,Guo2014,Liu2010,Nallapati2008-1,Nallapati2008-2}.
Influence can also come from outside the network.
Myers et al.~\cite{Myers2012} developed an information diffusion model to integrate both the internal influence and the external influence in a network. \looseness=-1

The major issue of hybrid methods is that they model influence at the document, user, or topic level.
In contrast, our method aims at detecting ideas at the word level and studying how the ideas correlate social groups and interact with each other along time.\looseness=-1





\section{Conclusions and Future Work}
In this paper,
we \doc{have aimed} to help users understand how ideas propagate between different groups.
To this end, we first derive an augmented bipartite word graph based on the correlations between words.
The major feature of the augmented bipartite graph is that its edge is represented by two vectors: a correlation vector and a lead-lag vector.
The two vectors, \doc{along} with the augmented bipartite graph, are formulated as a tensor representation.
Next, by factorizing the tensor, we automatically discover ideas that are represented by clusters of words.
For a certain pair of ideas, we further apply tensor factorization to cluster the time points, which can be used to segment the time series to identify the lead and lag period between ideas.
Finally, our evaluation demonstrates that the developed approach is generally more effective than the baseline methods.

One interesting direction for future work is exploring a wider range of text data representations for better tracking idea flows on social media.
Text data is typically treated as bag-of-words.
In this paper, we follow this paradigm since this representation is simple and easy to be understood.
There are some advanced techniques that can convert the unstructured text to a network with typed entity and relation information \cite{wang2015incorporating,Wang2015knowsim,Wang2016}, which can be less ambiguous than only using bag-of-words.
Integrating these techniques with our model is a very interesting topic for further study in the future.
In addition, we would like to study the lead-lag relationships within each user group and combine them with the lead-lag relationships between different user groups for better modeling idea flows on social media.

\section{Acknowledgements}
We would like to thank Mengchen Liu for valuable contributions on related work and the model formulation.
This research was supported by the National Key Technologies R\&D Program of China (2015BAF23B03) and a Microsoft Research Fund (No. FY15-RES-OPP-112).

\bibliographystyle{aaai}
\small
\bibliography{reference}
\end{document}